\documentclass[a4paper,preprint,showpacs,amssymb,pre,superscriptaddress]{revtex4}

\usepackage{graphicx}
\usepackage{graphicx}
\usepackage{amsfonts}
\usepackage{amssymb}
\begin{document}

\title{Resonant phenomena in extended chaotic systems
subject to external noise: the Lorenz'96 model case}

\author{Jorge A. Revelli}
\email{revelli@ifca.unican.es}
\author{Miguel A. Rodriguez}
\email{rodrigma@ifca.unican.es}
\author{Horacio S. Wio}
\email{wio@ifca.unican.es}

\affiliation{Instituto de F\'{\i}sica de Cantabria, Universidad de
Cantabria and CSIC, E-39005 Santander, Spain}

\begin{abstract}
We investigate the effects of a time-correlated noise on an extended
chaotic system. The chosen model is the Lorenz'96, a kind of ``toy"
model used for climate studies. Through the analysis of the system's
time evolution and its time and space correlations, we have obtained
numerical evidence for two stochastic resonance-like behavior. Such
behavior is seen when both, the usual and a generalized
signal-to-noise ratio function are depicted as a function of the
external noise intensity or the system size. The underlying
mechanism seems to be associated to a \textit{noise-induced chaos
reduction}. The possible relevance of these and other findings for
an \textit{optimal} climate prediction are discussed.
\end{abstract}

%\pacs{02.50.Ey, 05.40.-a, 05.45.-a}

\maketitle

\section{Introduction}

The last decades have witnessed a growing interest in the study of
the effect of noise on dynamical systems. It was proved that, under
some conditions, when a nonlinear dynamical system is subject to
noise, new phenomena can arise, phenomena that only occur under the
effect of such noise. All these phenomena are lump together under
the name \textit{noise-induced phenomena}. A few examples are:
\textit{stochastic resonance} in zero-dimensional and extended
systems \cite{RMP,extend1,extend2}, noise-induced transitions
\cite{lefev}, noise-induced phase transitions \cite{nipt1,nipt2},
noise-induced transport \cite{Ratch2,nipt3}, noise-sustained
patterns \cite{ga93,ber}, noise-induced limit cycle \cite{mangwio}.

Clearly, some of the above indicated noise-induced phenomena occur
in spatially extended systems, where another phenomena of great
relevance exists: \textit{spatio-temporal chaos} \cite{chaos00}.
However, studies on the effect of noise on spatially extended
chaotic systems are scarce \cite{chaos01}. There are studies on
chaotic systems where the \textit{pseudo-random} behavior of the
system is the trigger of phenomena usually associated with the
effect of a \textit{real} stochastic process (see for instance
\cite{rus,deter}). Hence, we can refer to the presence of a
\textit{deterministic} noise, that is a pseudo-noisy behavior
associated to the chaotic character of the system.

Among others, one of the most relevant and largely analyzed
application of studies of chaos in extended systems corresponds to
climate prediction. These kind of problems have been described by
Lorenz \cite{Lor96-1} as falling into two categories. On one hand
those which depend on the initial conditions, while on the other are
those depending on the boundaries. However, both kind of prediction
problems are affected by errors in the model equations used to
approximately described the behavior of real systems.

In a recent work \cite{Lor96-3}, and with the aim to improve the
limited weather predictability that results from a combination of
initial conditions uncertainty and model error, it was presented a
study of the effect of a stochastic parametrization in the Lorenz'96
model \cite{Lor96-1p,Lor96-2}. It is well known that much of the
current error in weather predictability derives from the practice of
representing the effects of process occurring at unresolved scales
by using simple forms of deterministic \textit{parametrization},
attempting to summarize the effects of small-scale processes in
terms of larger-scale, resolved, prognostic variables.

In this work, and with a similar objective as in a previously
indicated paper \cite{Lor96-3}, we investigate the effect of a
time-correlated noise on an extended chaotic system, analyzing the
competence between the above indicated \textit{deterministic} noise
and a \textit{real} stochastic process. In order to perform such a
study we have chosen the Lorenz'96 model \cite{Lor96-1}. In spite of
the fact that it is a kind of \textit{toy-model}, at variance with
the cases studied in \cite{chaos01}, it has a clear contact with
real systems as it is of interest for the analysis of climate
behavior and weather prediction \cite{Lor96-1p,Lor96-2,Lor96-3}. In
fact, this model has been heuristically formulated as the simplest
way to take into account certain properties of global atmospheric
models. To reach our objective, we have assumed that the only model
parameter is time dependent and composed of two parts, a constant
deterministic contribution plus a stochastic one.

Through the analysis of the system's temporal evolution and its time
and space correlations, we have obtained numerical evidence for two
\textit{stochastic resonance}-like (SR) \cite{RMP} behaviors. Such
behaviors are seen when both, the usual \textit{signal-to-noise
ratio} (SNR) and a \textit{generalized} function $SNR_{glob}$ (that
we call \textit{global} SNR), are depicted as function of the
external noise intensity or the system size. In accord with what was
shown in previous works \cite{RMP}, a SR phenomenon can occur in
systems without external periodic forcing, but having an internal
typical frequency. Hence, it seems reasonable to assume that the
present resonances typically occur at frequencies corresponding to a
system's internal quasi-periodic behavior, as well as at an
\textit{optimal} system's size. Finally, we discuss the possible
relevance of these findings for climate prediction.

\section{\label{model}The Model and Response Measures}

\subsection{The Model Lorenz'96}

The equations corresponding to the Lorenz'96 model
\cite{Lor96-1,Lor96-1p} are
\begin{eqnarray} \label{lor1}
\dot{x}_{j}(t) & = & - x_{j-1} (x_{j-2} - x_{j+1}) - x_{j} +
F. \\
j & = & 1,2,3,\ldots , N, \nonumber
\end{eqnarray}
where $\dot{x}_{j}(t)$ indicates the time derivative of $x_{j}(t)$.
In order to simulate a scalar meteorological quantity extended
around a latitude circle, we consider periodic boundary conditions:
$x_{0} = x_{N}, \,\,\,\, x_{-1} = x_{N-1}$.

As indicated before, the Lorenz'96 model \cite{Lor96-1} has been
heuristically formulated as the simplest way to take into account
certain properties of global atmospheric models. The terms included
in the equation intend to simulate advection, dissipation and
forcing respectively. In contrast with other toy models used in the
analysis of extended chaotic systems and based on coupled map
lattices, the Lorenz'96 system exhibits extended chaos ($F>9/8$),
with a spatial structure in the form of moving waves \cite{Lor96-1}.
The length of these waves is close to 5 spatial units. It is worth
noting that the system has scaled variables with unit coefficients,
hence the time unit is the dissipative decay time. In these units
the group velocity of the waves is close to $v_{gr} = 1.20$ implying
a eastward propagation. If, as done by Lorenz, we associate the time
unit to 5 days and the system size of 40 to the length of a latitude
circle, we have a highly illustrative representation of a global
model. If in addition we adjust the value of the parameter $F$ to
give a reasonable signal to noise ratio (Lorenz considered $F=8$)
the model could be most adequate to perform basic studies of
predictability. Hence, within this framework, the signal analyzed in
this paper would correspond to the passing of waves in a generic
observational site, in what is a simple mimic of forecasting at an
intermediate time range.

\subsection{Stochastic contribution}

As indicated before, here we assume that the model parameter $F$
becomes time dependent, and has two contributions, a constant and a
random one
\begin{equation} \label{lor0}
F_{j}(t) = F_{med} + \Psi_{j} (t),
\end{equation}
with $\Psi_{j} (t)$ a dichotomic process. That is, $\Psi_{j} (t)$
adopts the values $\pm \Delta$, with a transition rate $\gamma$:
each state ($\pm \Delta$) changes according to the waiting time
distribution $\psi_{i}(t) \sim e^{- \gamma t}$. The noise intensity
for this process is defined through \cite{vK,Gar} $\xi =\frac{\Delta
^2}{2\,\gamma }$.

\subsection{System Response}

As a measure of the SR system's response we have used the
\textit{signal-to-noise ratio} (SNR) \cite{RMP}. To obtain the SNR
we need to previously evaluate $S(\omega)$, the power spectral
density (psd), defined as the Fourier transform of the correlation
function \cite{vK,Gar}
\begin{equation} \label{lor3}
S (\omega) = \int _{-\infty}^{\infty} e^{i\omega \tau} \langle
x_{j}(0) x_{j}(\tau) \rangle \, d\tau ,
\end{equation}
where $\langle \,\,\, \rangle$ indicates the average over
realizations. As we have periodic boundary conditions simulating a
closed system, $\langle x_{j}(0) x_{j}(\tau) \rangle$ has a
homogeneous spatial behavior. Hence, it is enough to analyze the
response in a single site.

We consider two forms of SNR. In one hand the usual SNR measure at
the resonant frequency $\omega _{o}$ (that is, in fact, at the
frequency associated to the highest peak in $S(\omega)$) is
\begin{equation} \label{lor4}
SNR = \frac{\int ^{\omega _{o} + \sigma}_{\omega _{o} - \sigma} d
\varpi S(\varpi)}{\int ^{\omega _{o} + \sigma}_{\omega _{o} -
\sigma} d  \varpi S_{back}(\varpi)},
\end{equation}
where $2 \sigma$ is a very small range around the resonant frequency
$\omega _{o}$, and $S_{back}(\omega)$ corresponds to the background
psd. On the other hand we consider a global form of the SNR
($SNR_{glob}$) defined through
\begin{equation} \label{lor5}
SNR_{glob} = \frac{\int ^{\omega_{max}}_{\omega_{min}} d \omega
S(\omega)}{\int ^{\omega_{max}}_{\omega_{min}} d \omega
S_{back}(\omega)},
\end{equation}
where $\omega_{min}$ and $\omega_{max}$ define the frequency range
where $S(\omega)$ has a reach peak structure (with several
\textit{resonant} frequencies).

\section{\label{resu}Results}

We have analyzed the typical behavior of trajectories as $x_1(t) -
x_{med-T}$, where $x_{med-T}$ is the time average. It is worth
commenting that when the Lorenz´96 system evolves without external
noise (that is $F_j(t)$ is constant), the time evolution shows a
random-like behavior, with the main feature that the amplitude of
the oscillator is constant over all the time. However, when the
system is subject to a random force as described in Eq.
(\ref{lor0}), the temporal response decays, due to the fact that the
interaction between the intrinsic evolution and the external noise
produces a dissipative contribution on the system. Hence the
system's time evolution consists of a transitory regime and a
stationary one. This was analyzed through the behavior of the
``decay" of $\langle x_1 (t) - x_{med-T}\rangle$. We assumed that
this decay can be adjusted by an exponential law. The decay
parameter ($\lambda$) only depends on $F_{med}$ and it does not
depend neither on the system size nor on the noise intensity. This
analysis is relevant when studying the effects of noise on the
stationary regime. From those results it was possible to anticipate,
and approximately identify, the existence of two regimes: a weak or
undeveloped chaos for $F_{med} < 6.0$, and strong or completely
developed chaos for $F_{med} > 6.0$.

The typical numbers we have used in our simulations are: averages
over $10^3$ histories, and $\sim 10^4$ simulation time steps (within
the stationary regime, see later).

\begin{figure}
\centering
\resizebox{.48\columnwidth}{!}{\includegraphics{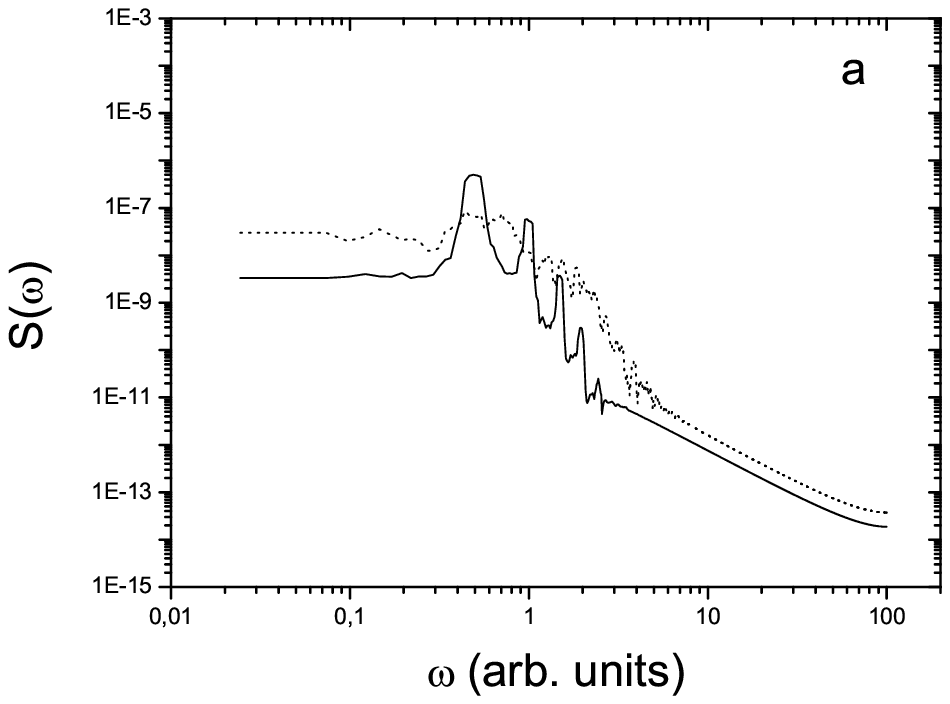}}
\resizebox{.48\columnwidth}{!}{\includegraphics{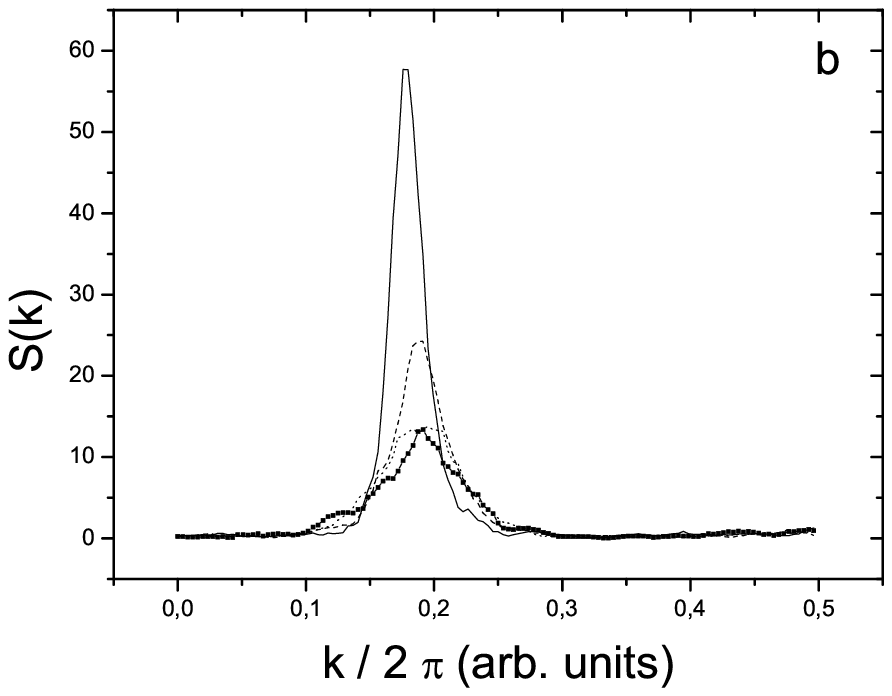}}
\caption{(a) $S(\omega)$ for a couple of values of $F_{med}$
($F_{med} = 4.5, 7.8$) and for a small noise intensity ($\xi = 5
\times 10^{-4}$, $\Delta = 0.1$ and $N=128$). (b) Spatial spectrum
for $N=256$, $\Delta = 0.001$ and $\gamma = 10$, for several values
of $F_{med}$: continuous line $F_{med}=5$, dashed $F_{med}=6$,
dotted $F_{med}=7$ and dash-dotted $F_{med}=8$ .} \label{Fig.1}
\end{figure}

We have evaluated the psd $S(\omega)$ in a standard way. Figure
\ref{Fig.1}-a shows the typical form of the psd $S(\omega)$ for a
couple of values of $F_{med}$ ($F_{med} = 4.5, 7.8$) and for a noise
intensity $\Delta = 0.1$ ($\xi = 5 \times 10^{-4}$). The figure
shows a rich peak structure within the interval $0.22 < \omega <
1.3$. It is worth to comment that the frequencies associated to the
different peaks seems to correspond to the harmonics of the main (or
first) peak frequency. In Fig. \ref{Fig.1}-b, we show the form of
$S(k)$, the associated spatial spectrum. Here we depict the spectrum
for fixed values of the system's size ($N=256$), and noise intensity
($\Delta = 0.001$ and $\gamma = 10$), and different values of
$F_{med}$. The independence of the position of the peak (indicating
a single spatial structure of wavelength $k/2\pi=0.2$) is apparent.
However there is a strong dependence on the peak intensity when
varying $F_{med}$, from a net peak for underdeveloped chaos
($F_{med}=5$) to a reduced peak for well developed chaos
($F_{med}=8$). It is worth here remarking that there is no
dependence (or eventually a very weak one) of this behavior on the
noise intensity.

\begin{figure}
\centering
\resizebox{.6\columnwidth}{!}{\includegraphics{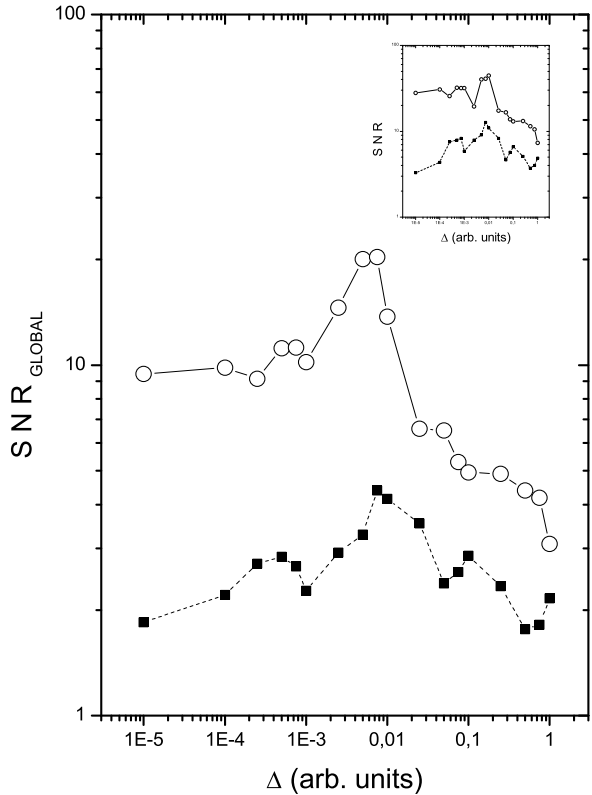}}
%\resizebox{.45\columnwidth}{!}{\includegraphics{Fig-06.jpg}}
\caption{$SNR_{glob}$ vs $\Delta$, for fixed values of $N=64$ and
$\gamma = 10$; and a couple of values of $F_{med}$: white circles
$F_{med}=5$, black squares $F_{med}=6$. The insert shows the
behavior of the usual $SNR$ for the same cases.}
\label{Fig2}
\end{figure}

Figure \ref{Fig2} shows the dependence of $SNR_{glob}$ --for a
space-temporal noise--  on $\Delta$ for fixed values of $N$ and a
couple of values of $F_{med}$. It shows a peak for $\Delta \sim 6-7
\times 10^{-3}$, that corresponds to the \textit{fingerprint} of the
more usual form of SR. The insert shows the same case but for $SNR$.

The analysis of the dependence of $SNR_{glob}$ on $F_{med}$ have
also shown the existence of the previously indicated two regimes: a
weak or undeveloped chaos for $F_{med} < 6.0$, and a strong or
completely developed chaos for $F_{med} > 6.0$. Those regimes are
characterized by the existence of well defined peaks in the psd, in
the former case, and a less defined peak structure in the latter
case, as seen in Fig. \ref{Fig.1}a .

It is worth to detach the strong similarities in the behavior of
$SNR$ and $SNR_{glob}$ --which becomes apparent when comparing the
main Fig. \ref{Fig2} with its insert-- indicating that the second
one is an adequate and more versatile measure to characterize the
system's response. Hence, due to the clearness in the determination
of $SNR_{glob}$ (compared with the difficulties for a correct
determination of $SNR$ for large values of $F_{med}$) in what
follows we adopt it for the system's analysis.

\begin{figure}
\centering
\resizebox{.6\columnwidth}{!}{\includegraphics{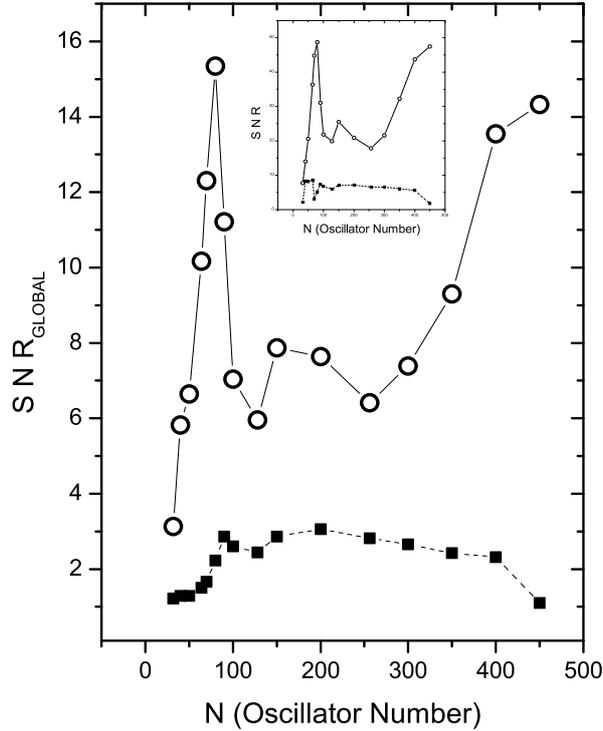}}
\caption{$SNR_{glob}$ vs $N$, for fixed values of $\Delta =0.1$ and
$\gamma = 5$ ($\xi=0.001$), and a couple of values of $F_{med}$:
white circles $F_{med}=5$, black squares $F_{med}=6$. The insert
shows the behavior of the usual $SNR$ for the same cases.}
\label{Fig3}
\end{figure}

In Fig. \ref{Fig3} we depict the dependence of $SNR_{glob}$ on $N$,
for the case of space-temporal noise, for fixed values of $\Delta
=0.1$ and $\gamma = 5$ ($\xi=0.001$) and a couple of values of
$F_{med}$. The existence of the peak at $N \sim 60$ for $F_{med} =
5.0$ is apparent. In addition, we observe an increase of
$SNR_{glob}$ for large values of $N$. However, for $F_{med} = 6.0$,
the peak has disappeared, as well as the increase with larger values
of $N$.  The presence of the peak at $N \sim 60$ indicates a kind of
\textit{system-size stochastic-resonance} (SSSR) \cite{sssr}. The
insert shows the same case but for $SNR$. Again, as indicated above,
the nice agreement between the behavior of $SNR_{glob}$ and that of
$SNR$.

The figures clearly show that the system's response (SNR) is
stronger when the system is in the underdeveloped chaos range than
when it is in the highly-developed chaos one. Our results also show
that the main resonant frequency does not depend on the noise
intensity, system size, or correlation rate.

\begin{figure}
\centering
\resizebox{0.6\columnwidth}{!}{\includegraphics{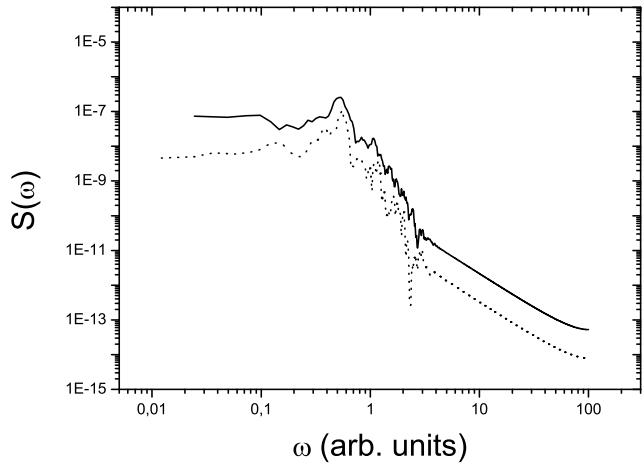}}
%\onefigure{paperfinal-5-f.EPS}
\caption{$S(\omega)$ with and without external noise, for $F=5$ and
$\gamma = 10$. Continuous line $\Delta=0$, dot-line $\Delta=0.1$.}
\label{Fig0}
\end{figure}

We want to close this section commenting that the SR phenomena found
here looks similar to the so called \textit{internal signal} SR
\cite{int}. In previous studies it was shown that in some systems
having an internal typical frequency, SR can occur not only at the
frequency of an external driving signal, but at the frequency
corresponding to the internal periodic behavior \cite{int}.
Regarding the present mechanism of SR, what we can indeed comment is
that the increase in the SNR is related not to a
\textit{reinforcement} of the peak high respect to the noisy
background at a given frequency, but with a \textit{reduction} of
the \textit{pseudo} (or deterministic) noisy background when turning
on the \textit{real} noise. That is, the interplay between ``real"
noise and ``deterministic" noise conforms a kind of
\textit{noise-induced chaos reduction}. Figure \ref{Fig0} shows, for
fixed values of $F$ and $\gamma$, the behavior of $S(\omega)$ in
both cases: with ($\Delta \neq 0$) and without noise ($\Delta = 0$).
The above indicated reduction trend, as the \textit{real} noise is
turned on, is apparent. However, the present mechanism is not
completely clear so far, and requires further studies.

\section{\label{conc}Conclusions}

We have investigated the effect of a time-correlated noise on an
extended chaotic system, analyzing the competence between the
indicated \textit{deterministic} or \textit{pseudo-noise} and the
real random process. For our study we have chosen the Lorenz'96
model \cite{Lor96-1} that, in spite of the fact that it is a kind of
\textit{toy} model, is of interest for the analysis of climate
behavior \cite{Lor96-2,Lor96-3}. It worth remarking that it accounts
in a simple way of the spatial structure of geostrophic waves and
the dynamics of tropical winds. The time series obtained at a
generic site $x_i(t)$ mimics the passing of such waves, which is in
fact a typical forecast event. We have assumed that the unique model
parameter $F$, is time dependent and composed of two parts, a
constant deterministic, and a stochastic contribution in a
spatial-temporal form.

We have done a thorough analysis of the system's temporal evolution
and its time and space correlations. From our results it is clear
that, using two complementary SNR measures, a usual and a global
one, we have obtained numerical evidence for two SR-like behaviors.
In one hand a ``normal" SR phenomenon occur at frequencies that seem
to correspond to a system's quasi-periodic behavior. On the other
hand, we have found a SSSR-like behavior, indicating that there is
an optimal system size for the analysis of the spatial system's
response. As indicated before, the effect of noise is stronger when
the chaos is underdeveloped.

We argue that these findings are of interest for an \textit{optimal}
climate prediction. It is clear that the inclusion of the effect of
an external noise, that is a stochastic parametrization of unknown
external influences, could strongly affect the deterministic system
response, particularly through the possibility of an enhanced
system's response in the form of resonant-like behavior. It is worth
here remarking the excellent agreement between the resonant
frequencies and wave length found here, and the estimates of Lorenz
\cite{Lor96-1p}.

The effect of noise is weak respect to changes in the spatial
structure, with the main frequencies remaining unaltered, but it is
strong concerning the strength of the ``self-generated"
deterministic noise. In fact, in such a system and at the resonant
frequencies, forecasting could be improved by the external noise due
to the effect of suppression of the self-generated chaotic noise.
The detailed analysis of such an aspect will be the subject of a
forthcoming study \cite{nos1}.

\acknowledgments We acknowledge financial support from MEC, Spain,
through Grant No. CGL2004-02652/CLI. JAR thanks the MEC, Spain, for
the award of a \textit{Juan de la Cierva} fellowship. HSW thanks to
the European Commission for the award of a \textit{Marie Curie
Chair} during part of the development of this work.

\section*{Acknowledgements}

We acknowledge financial support from MEC, Spain, through Grant No.
CGL2004-02652/CLI. JAR thanks the MEC, Spain, for the award of a
\textit{Juan de la Cierva} fellowship. HSW thanks to the European
Commission for the award of a \textit{Marie Curie Chair} during part
of the development of this work.

\end{document}